\documentclass[preprint,12pt]{elsarticle}
\pdfoutput=1

\usepackage{amsmath,amssymb,color,caption}
\newcommand{\be}{\begin{equation}}
\newcommand{\ee}{\end{equation}}
\newcommand{\bel}[1]{\begin{equation}\label{#1}}
\newcommand{\beal}[1]{\begin{eqnarray}\label{#1}}
\newcommand{\bea}{\begin{eqnarray}}
\newcommand{\eea}{\end{eqnarray}}
\def\nn{\nonumber \\}
\def\error{\mathop{\mathrm{error}}}
\def\bx{\mathbf{x}}
\def\by{\mathbf{y}}

\journal{Journal of Computational Physics}

\begin{document}

\begin{frontmatter}

\title{Auxiliary variables for nonlinear equations with softly broken symmetries}
\date{}
\author{Ken. D. Olum}
\ead{kdo@cosmos.tufts.edu}
\author{Ali Masoumi\corref{cor1}}
\cortext[cor1]{Corresponding author}
\ead{ali@cosmos.phy.tufts.edu}
\address{Institute of Cosmology, Department of Physics and Astronomy,
Tufts University, Medford, MA 02155, USA}

\begin{abstract}
General methods of solving equations deal with solving $N$ equations
in $N$ variables and the solutions are usually a set of discrete
values.  However, for problems with a softly broken symmetry these
methods often first find a point which would be a solution if the
symmetry were exact, and is thus an approximate solution.  After this,
the solver needs to move in the direction of the symmetry to find the
actual solution, but that can be very difficult if this direction is
not a straight line in the space of variables.  The solution can often
be found much more quickly by adding the generators of the softly
broken symmetry as auxiliary variables.  This makes the number of
variables more than the equations and hence there will be a family of
solutions, any one of which would be acceptable.  In this paper we
present a procedure for finding solutions in this case, and apply it
to several simple examples and an important problem in the physics of
false vacuum decay.  We also provide a Mathematica package that
implements Powell's hybrid method with the generalization to allow
more variables than equations.
\end{abstract}

\end{frontmatter}

\section{Introduction}
Numerical methods of solving equations are one of the most important
topics in numerical analysis. There is a plethora of techniques which
each is adequate for specific sets of problems. For examples refer to
\cite{NumericalRecipe} and references therein. All techniques attempt
to successively improve some guess for the variable values.

The best-known of these is Newton's method (also called
Newton-Raphson), which works by linearizing the equations around the
current guess and solving the linearized problem.  It converges very
rapidly when it gets near the solution, but if it is not close to the
solution it can take large steps in unhelpful directions that make the
next guess worse than the previous one.

An alternative is to attempt to minimize the sum of the squares of a
set of functions by descending the gradient of the total error.  This
method improves the solution at each step but can be very slow.
There are also hybrid methods which combine these two \cite{Powell1}.

The general techniques of solving equations involve solving a set of
$N$ equations of $N$ variables where the solution is usually a
discrete set of vectors. However, for problems with softly broken
symmetries, as described below, these techniques can be very slow,
sometimes so slow that the solution cannot be found.  The problem is
that there can be a complicated manifold of approximate solutions
related by the broken symmetry.  Once the technique finds one such
solution, almost every possible step will make the error much worse.
Any technique that depends on making progress in the sense of
decreasing the error will be forced to take a very large number of
tiny steps.

To solve such  problems we propose adding auxiliary variables
which are the generators of the softly broken symmetry, without adding
new equations. The idea of adding variables for solving equations is not new,
but it is usually accompanied by adding an equal number of equations. In our method we add new variables only.

This paper is organized as follows.  In Section~\ref{sec:formalism} we
lay out the formalism for adding new variables.  In
Section~\ref{sec:SimpleExamples} we present two simple examples with
softly broken symmetries and the reason for adding new unknowns. In
Section~\ref{sec:tunneling} we present a specific theoretical physics
problem that can be solved this way.  In Section~\ref{sec:package} we
briefly introduce a Mathematica package for Powell's hybrid method
and we conclude in Section~\ref{sec:Conclusion}

\section{Problem, formalism and algorithm} \label{sec:formalism}
Suppose one wants to solve a set of equations in the form 
\be
	f_i(\bx)=0~, \qquad i=1,\ldots, N~,
\ee
where $\bx$ has $N$ elements, using an equation solver, which could use
Newton, gradient or hybrid methods. This always starts by choosing an
initial guess for the solution $\bx_0$ and taking steps according to
some rule for approaching the solution. Of course the initial guess will
not be the solution and hence the functions $f_i(\bx_0)$'s do not
vanish. The goal of taking a step is to find a new set of points
$\bx_1, \bx_2, \ldots, \bx_I$ in such a way that the values of the
$f_i$ in each step get closer to zero, at least in some aggregate
sense.

Let us define the error as
\bel{error}
\error(\bx)=\left(\sum_{i=1}^N{f_i(\bx)^2}\right)^{1/2}~.
\ee
In some cases, for example Fig.~\ref{fig:RotManfiold} below, the
geometry the of error function is such that there are narrow valleys,
which makes the steps very small without much progress in each step.
In this case the error along the valley decreases very slowly because
the error function has an approximate rotational symmetry. If one can
add a variable which allows the solver move to quickly along the valley,
the generator of the rotation in the case of
Fig.~\ref{fig:RotManfiold}, the solution can be found much
more easily. Notice that this is a very simple version of the problem.
Real problems can have valleys which have sharp turns, and for more
than two variables the geometry can be much more complicated.

Suppose we now add extra variables $\by=\{y_1, y_2, \ldots, y_K\}$ and
extend our functions $f_i(\bx)$ to some $g_i(\bx, \by)$ such that
$g_i(\bx, 0)= f_i(\bx)$.  It is now possible to move in new directions
given by $\by$.  If these variables have been chosen well, the solver
will be able to take large steps in the new directions that quickly
reduce the error and lead to a solution.  In particular, if the
valley of approximate solutions is straight, or at least close
to straight, in the new variables, then the solver can make rapid
progress.  We show examples of this phenomenon in
Sections~\ref{sec:SimpleExamples} and \ref{sec:tunneling}.

Of course, since we add variables without adding constraints, the
solution is not unique.  For each solution of $f_i(\bx)=0$ there will
be a $K$-dimensional family of solutions to $g_i(\bx, \by)=0$.  When
the process succeeds, we have some solution $\{\bx, \by\}$, and we
need to get from there to the solution of the original problem
$f_i(\bx)=0$.  But this is straightforward if the $\by$ are the
generators of a symmetry.  We can solve the original problem by simply
applying the symmetry transformation to the $\bx$ that we found.

We now present an algorithm for finding a solution of $N$ equations of
$N+K$ variables using Powell hybrid techniques.  Powell's hybrid (or
``dogleg'') method \cite{Powell1} takes steps which are a combination
of the step recommended by Newton's method and a step in the direction
that most rapidly reduces $\error(\bx)$, i.e., the negative of the
gradient of $\error(\bx)$.  In the gradient case, the procedure is
unaffected by additional variables.  We simply have a scalar function
of $N+K$ variables whose gradient we descend.

The application of Newton's method is slightly more complicated.
Newton's method consists of linearizing the equations around some
point $\bx$ and solving the linearized equations.  With $N$ equations
in $N+K$ variables, there will be a $K$-dimensional subspace of
solutions.  In this case, we choose the solution which is nearest to
the current guess in the Euclidean metric on the $N+K$ variables.
This is straightforwardly determined by singular value decomposition
of the rectangular Jacobian.

The steps chosen by gradient methods depend on the scaling of the
variables (and also of the functions).  Such methods try to find small
steps in the space of variables that will lead to large improvements
in the error.  This feature can be used to control how much the solver
tries to make use of the additional variables.  If an additional
variable is multiplied by some factor $s>1$ before being used in the
function evaluation, the solver will give higher weight by factor $s$
to changes produced by this variable, and will be thus more likely to
exploit that additional variable to simplify the problem.

The unmodified Newton's method is invariant under rescaling of the
variables (and the functions), because that doesn't change the
solution to the linearized problem.  However, the modified method here
chooses the closest point in the subspace of solutions.  The factor
$s$ increases the effect of the additional variable, so the change in
it to achieve the same result is smaller.  Thus increasing $s$ causes
the modified Newton's method to choose a solution whose differences
are more due to the affected variable and less to the others.
\section{Simple examples}\label{sec:SimpleExamples}
In this section we present two problems with a softly broken symmetry  and show the progress made by adding a new variable. 
\subsection{Softly broken rotational symmetry}\label{sec:rotationalSym}
\begin{figure}
   \centering
   \includegraphics[width=2.5in]{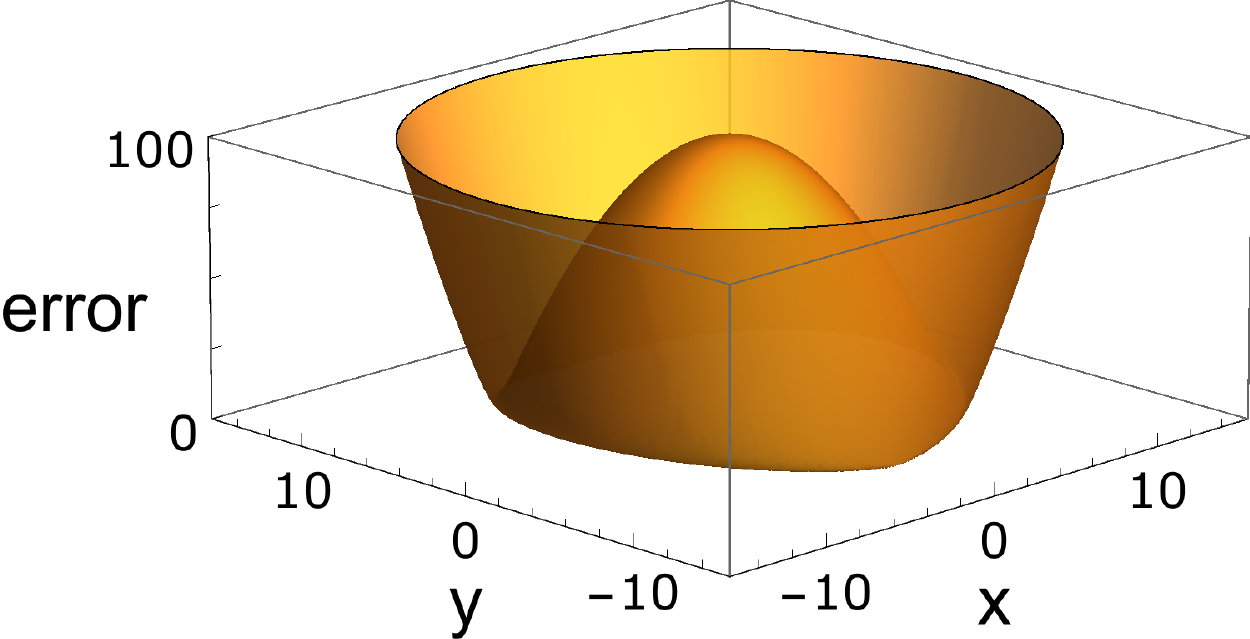}
   \includegraphics[width=2.5in]{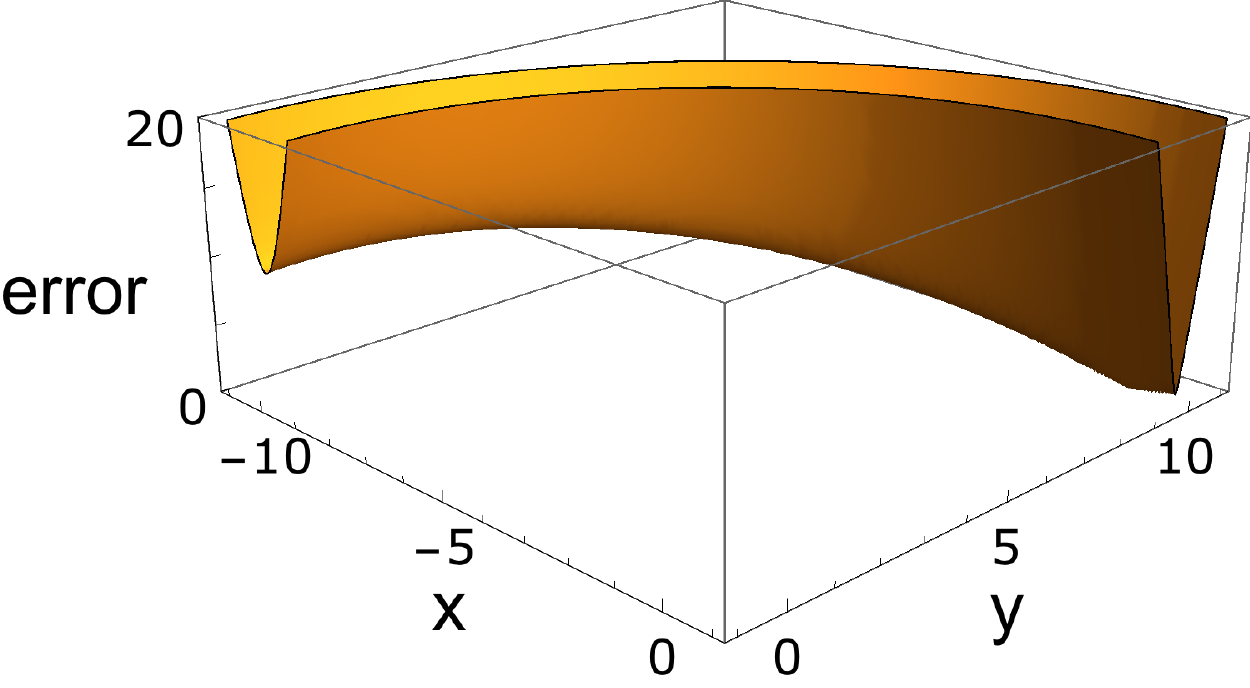}
   \includegraphics[width=1.5in]{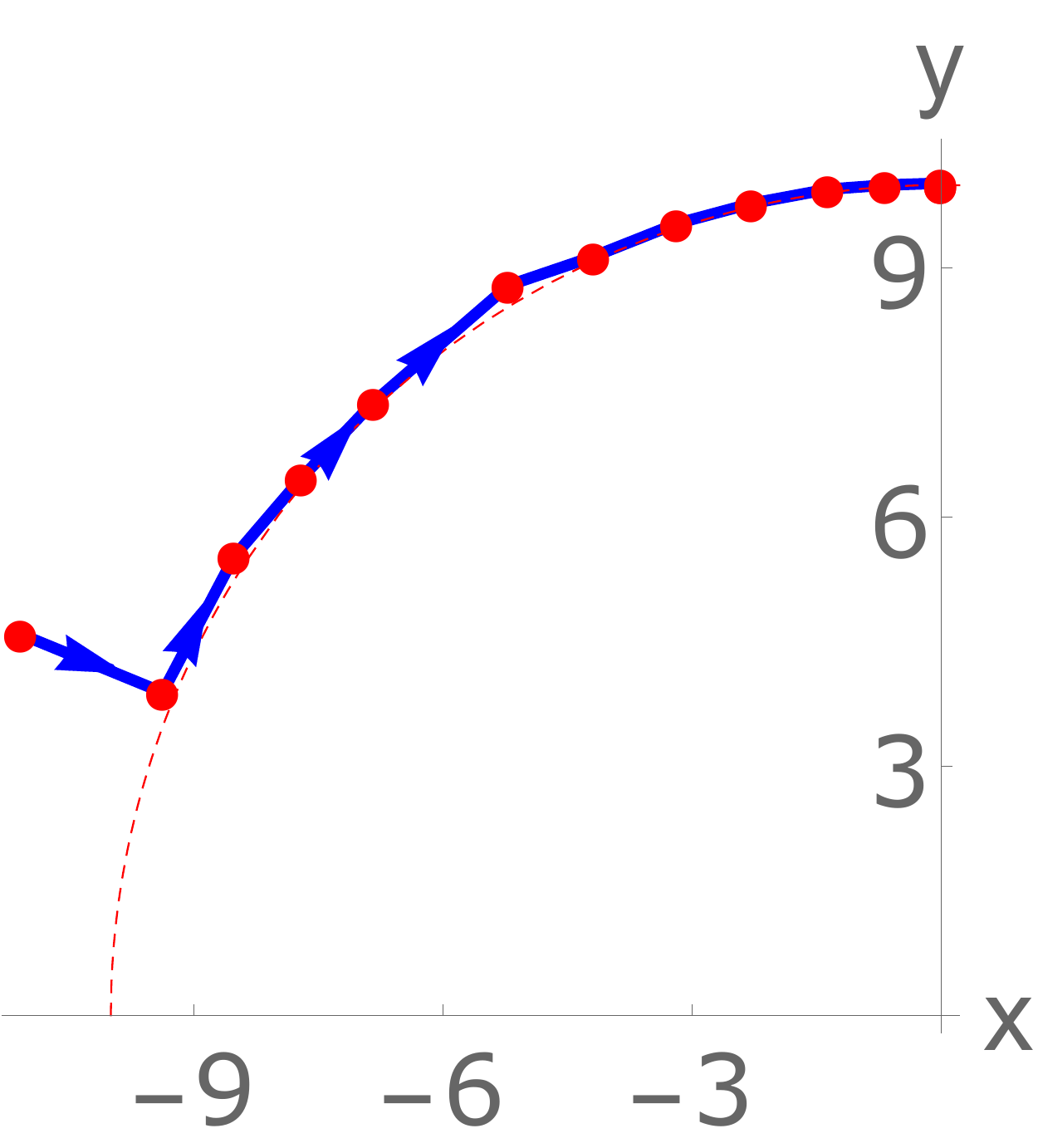}
   \includegraphics[width=2.8in]{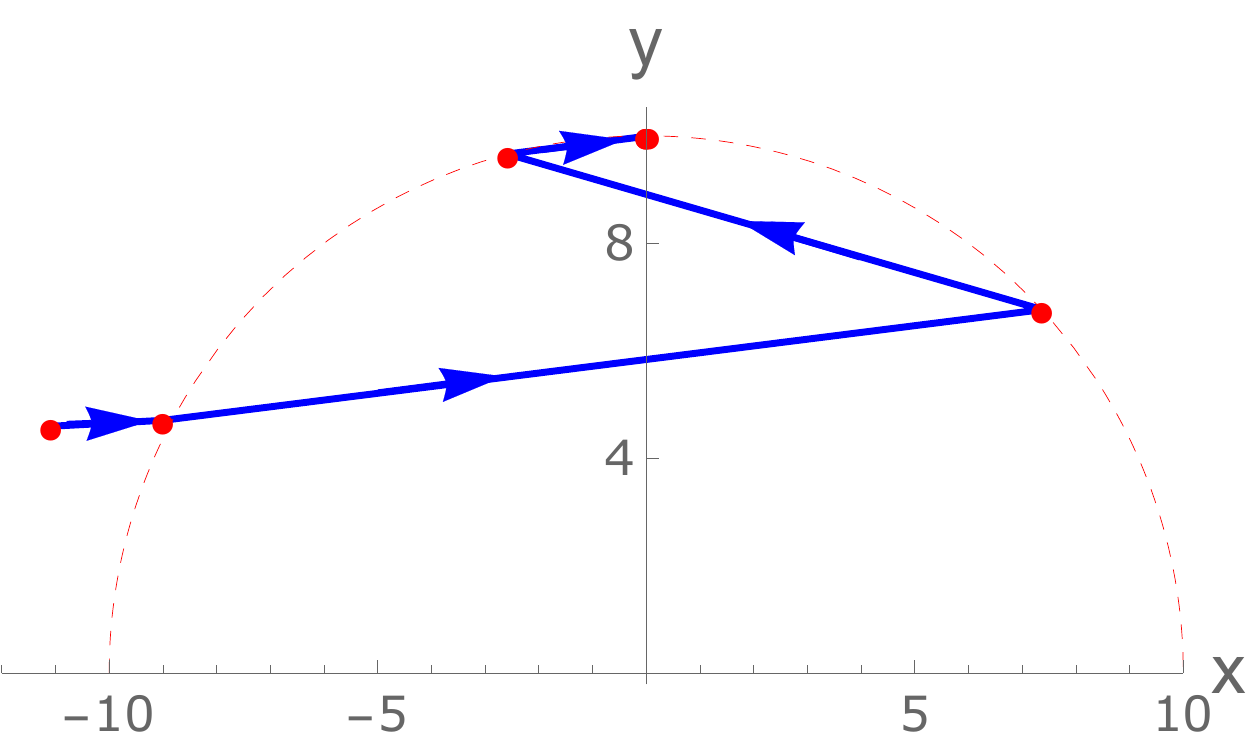}
   \caption{ Top, the error as a function of $x$ and $y$ for the
     functions defined in \eqref{rotSym} with $a=10$ and $b=1$. The
     problem has an approximate rotational symmetry. To see the
     breakdown of the symmetry on the right panel we zoomed in on the
     valley, which is slanted.  The hybrid solver takes many small
     steps before finding the solution. On the bottom left panel we
     show the path that the hybrid solver took without an extra
     variable.  On the right we show the path in $x$ and $y$ taken
     when we use the extra variable $\theta$.  In this case, the
     solver takes steps in $x'$, $y'$, and $\theta$, but we graph the
     points $x$ and $y$ given by \eqref{rotChangeOfVar}. In both cases
     the starting guess is $(-12 \cos(\pi/8), 12 \sin(\pi/8))$. The
     left panel took 14 steps and the right only 5 steps. We see in
     the next figure that smaller $b$ increases the number of steps
     rapidly. }
\label{fig:RotManfiold}
\end{figure}

We start from a very easy problem to demonstrate the main idea behind
our proposal.  Consider the functions
\beal{rotSym}
	f_1(x, y)&=& x^2+y^2 -a^2~, \nn
	f_2(x,y)&=&b x~,
\eea
for some parameters $a$ and $b$.  We want to find $x$ and $y$, such
that $f_1 = f_2 = 0$.  The solution is $y=\pm a$ and $x=0$, but we
suppose we don't know that and are trying to find the solution
numerically.
In the top panel of Fig.~\ref {fig:RotManfiold}, we plot the 
error for this set of equations with $a=10$ and $b=1$, as a function
of the variables $x$ and $y$. The upper right panel shows a magnified
picture of the curving and slanted valley.  In the lower left of
Fig.~\ref {fig:RotManfiold}, we show the 14 steps taken by Powell's
method to find the solution.

Now let us introduce an extra variable which allows
rotation, the broken symmetry.  Instead of using $x$ and $y$ and
variables, we use $x'$, $y'$, and $\theta$, with $x$ and $y$ given by
\beal{rotChangeOfVar}
	x &=& x' \cos \theta + y' \sin \theta~, \nn
	y&=& -x' \sin \theta+ y' \cos \theta~.
\eea
Our functions thus become
\beal{rotSym2}
	f_1(x, y)&=& x'^2+y'^2 -a^2~, \nn
	f_2(x,y)&=& b (x' \cos \theta + y' \sin \theta)~.
\eea
Now there will be an infinite number of solutions.  For each value of
$\theta$, there will be a solution $x'$ and $y'$. We can recover the
values of $x$ and $y$ using \eqref{rotChangeOfVar}. As shown in the
lower right panel of Fig.~\ref{fig:RotManfiold} this problem is now solved
in 5 steps.

\begin{figure}
   \centering
   \includegraphics[width=1.7in]{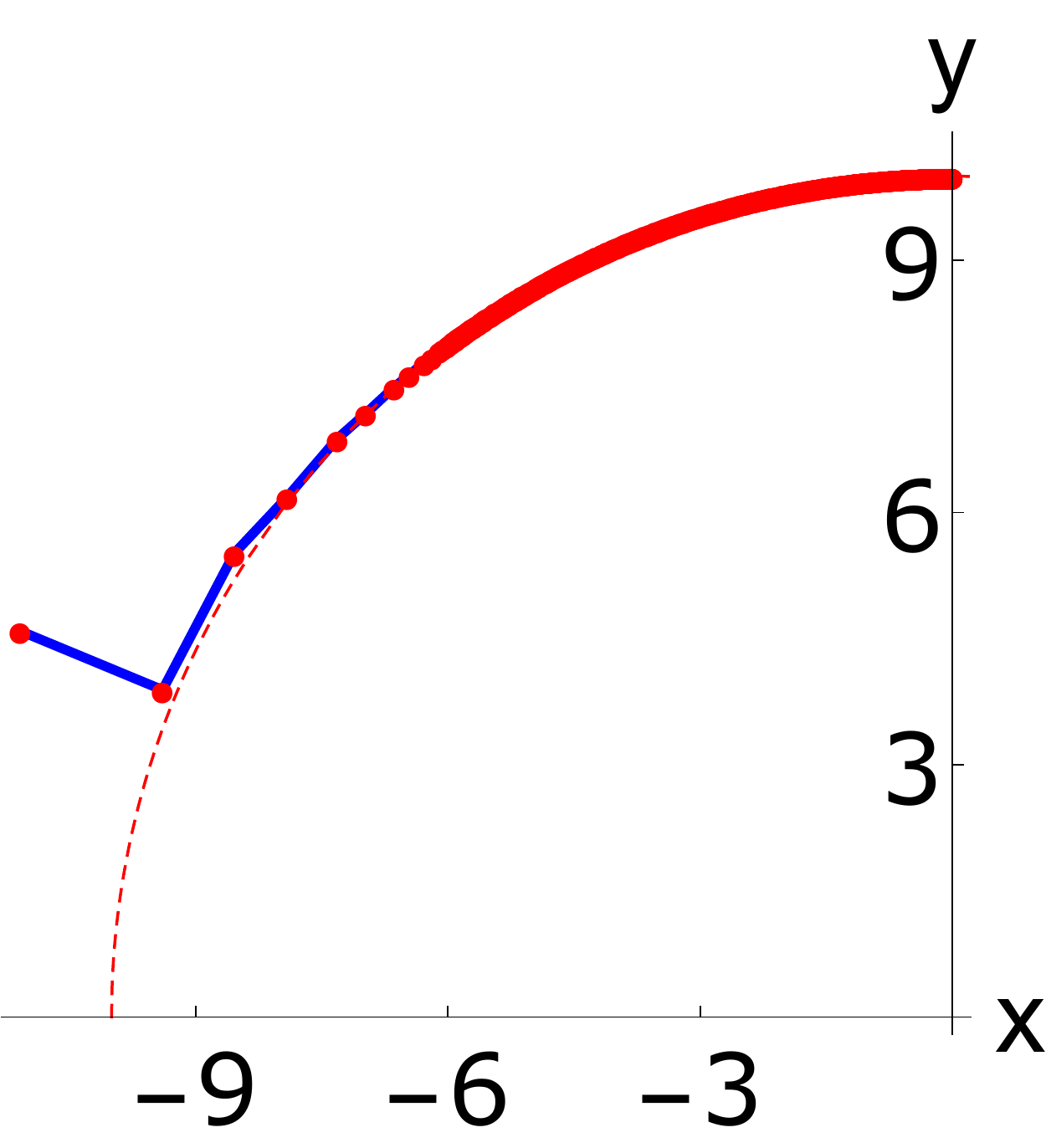}
   \includegraphics[width=3.2in]{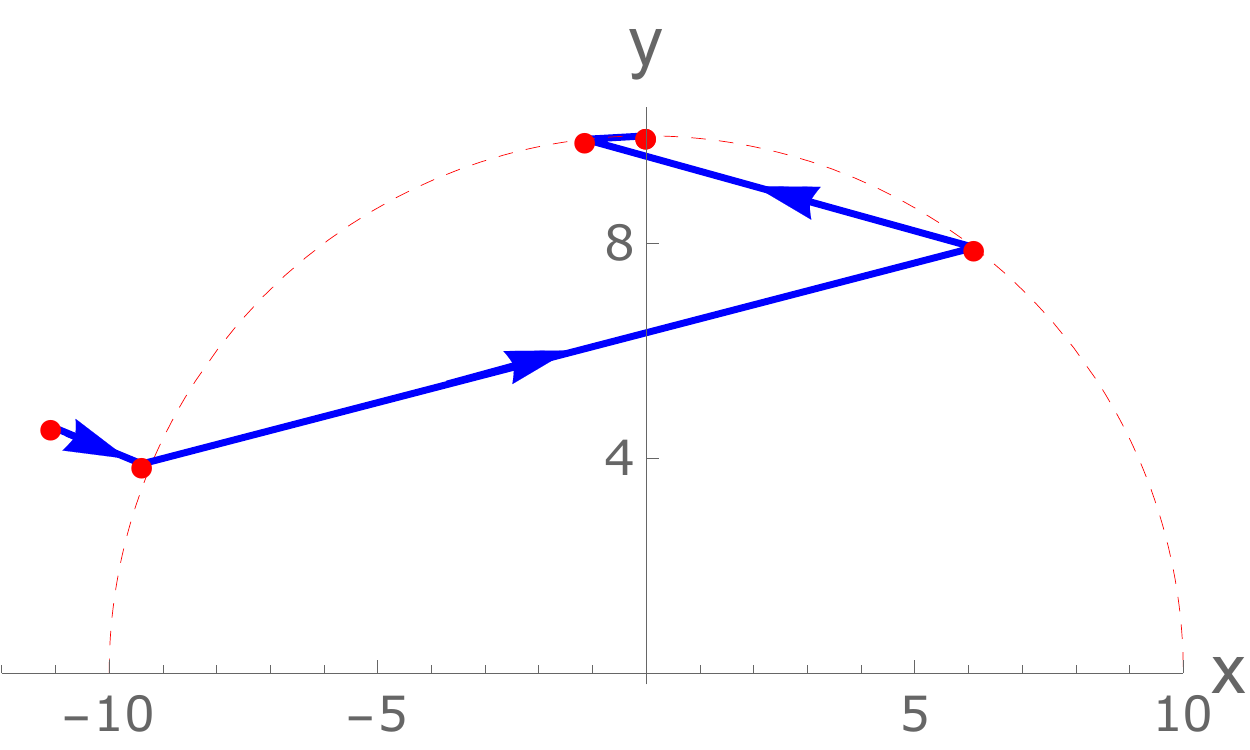}
   \caption{ The steps taken by the solver for the example presented
     in \eqref{rotSym} for $a=10$ and $b=10^{-3}$. Now the rotational
     symmetry is broken very softly. Left shows the steps taken without
     the extra variable and right with an extra variable that
     generates the broken symmetry. In both cases the starting guess
     is $( -12\cos(\pi/8), 12 \sin(\pi/8))$. The left panel took 537
     steps and the right only 5 steps.  }
   \label{fig:RotManfiold2}
\end{figure}

In Fig.~\ref{fig:RotManfiold2}
we show the path our solver took for $a=10$ and $b=10^{-3}$. The
symmetry here is broken softly, so there is not much of a slant in the
valley. If one uses a hybrid or gradient solver it takes a large
number of steps to find the solution.  For example Powell's
method took 537 steps, mostly creeping slowly along the valley, whereas
with $\theta$ the solution can still be found in 5 steps.

This particular problem is rather trivial.  Instead of adding a
variable one could simply work in polar coordinates, which manifest
the symmetry, and get the solution right away. This is possible
because one can easily parameterize the remaining degree of freedom
after the symmetry has been factored out.

One also can solve this problem by using a pure Newton method.
Since one of the equations is linear, Newton's method will solve it
exactly in every step by jumping to some point on the $y$ axis.
Newton's method does not care about rescaling the equations and hence
the smallness of $b$ is not important and the symmetry is not broken
softly.

In the next sections we present two other problems where there is no
obvious change of coordinate and the Newton technique does not
necessarily work well.

\subsection{Softly broken translational symmetry}\label{sec:translationalSym}
In this section  we present a problem where the correct
parameterization is not as trivial as the previous one.  Suppose we
have some functions $f(x)$ and $g(x)$, chosen from families of similar
functions, and we consider the function given by
\bel{trans1}
	F(x)= \left\{
		\begin{array}{ll}
			f(x) & f(x)<0~, \\
			g(x) & g(x)\ge0~.
		\end{array}\right. 
\ee
Here $f(x)$ and $g(x)$ are monotonically increasing functions. 
That to say that $F$ follows $f$ from $x=-\infty$ until $f(x)$ reaches
0, and $g$ from $x=\infty$ toward smaller $x$ until $g(x)$ reaches 0.
To have a well-defined function, we would like $f(x)$ and $g(x)$ to
reach 0 at the same point, and to have a $C^1$ function we would like
the derivatives of $f$ and $g$ to match also at this point.

Our $f$ and $g$ will be chosen from the following classes of
functions,
\beal{trans2}
	f(x) &=& a e^ x  + \epsilon e^{2x} - 1~,\nn
	g(x) &=& 1 - b e^{-(1+\epsilon) x}~.
\eea
where $a$ and $b$ specify which functions we chose and
$\epsilon$ is a fixed small parameter.

To have a problem where the symmetry is easily shown, rather than
specifying $a$ and $b$ we will specify that $F$ takes on a given value
$F_1<0$ at a point $x_1$ and similarly a given value $F_2>0$ at a
given point $x_2$.  We will then attempt to vary $F_1$ and $F_2$ to
find a well-defined $C^1$ function F.

First we explain the broken symmetry. If $\epsilon=0$, we have
\bea
	f(x)&=& a e^ x  - 1\\
	g(x)&=& 1 - b e^{-x}~.
\eea
We want the two functions to join (vanish) at some $x_0$. For each value $x_0$
in the interval $(x_1,x_2)$ there will be a solution in the form
\bea
	f(x_1)&=& e^{x_1-x_0}-1~, \\
	g(x_2)&=&1-e^{x_0-x_2}~.
\eea
Hence the values of $F_1$ and $F_2$ only depend on $x_1-x_2$
and are invariant under a shift in these two numbers. 

The terms which include $\epsilon$ break this symmetry softly and
there will be unique values for $F_1$ and $F_2$ which make the
function smooth. But because $\epsilon$ is small, it is difficult to
find the correct values once we find some which solve the $\epsilon=0$
problem.

\begin{figure}
   \centering
   \includegraphics[width=2.5in]{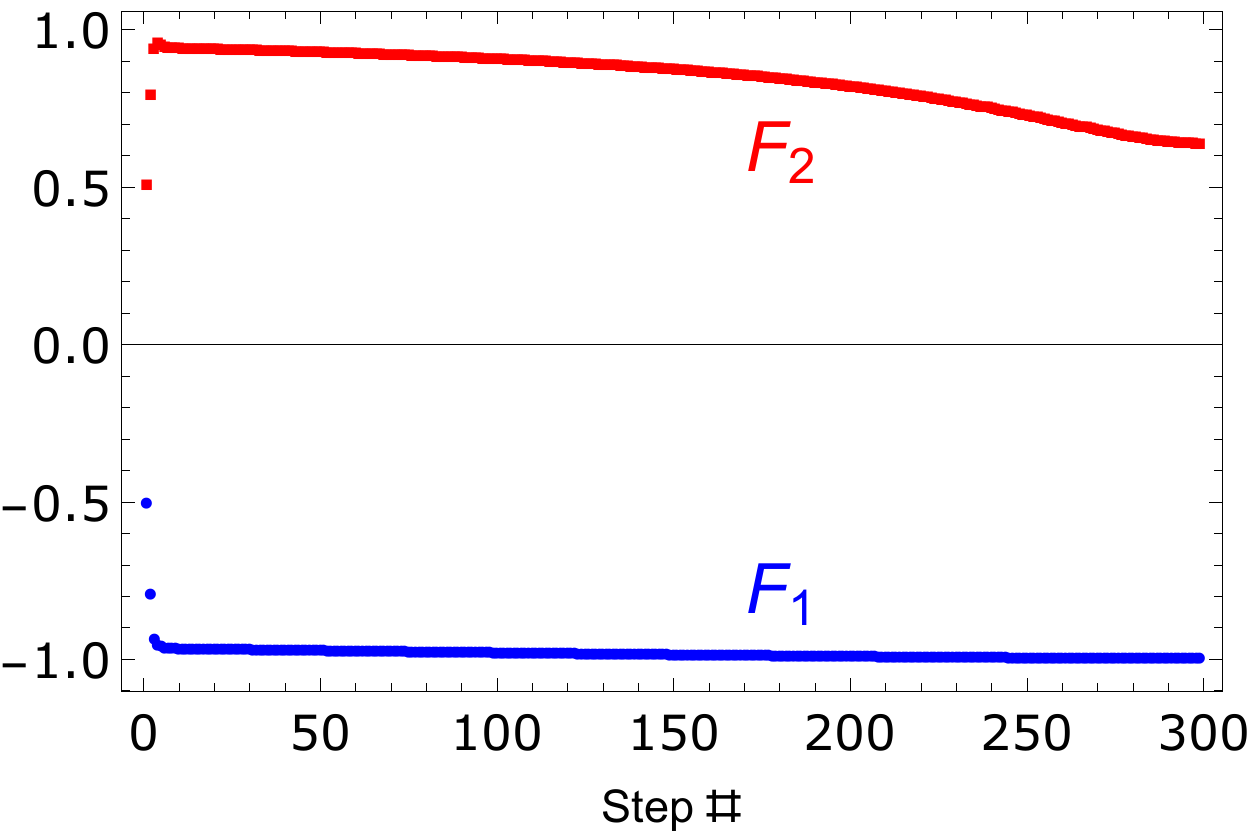}
   \includegraphics[width=2.5in]{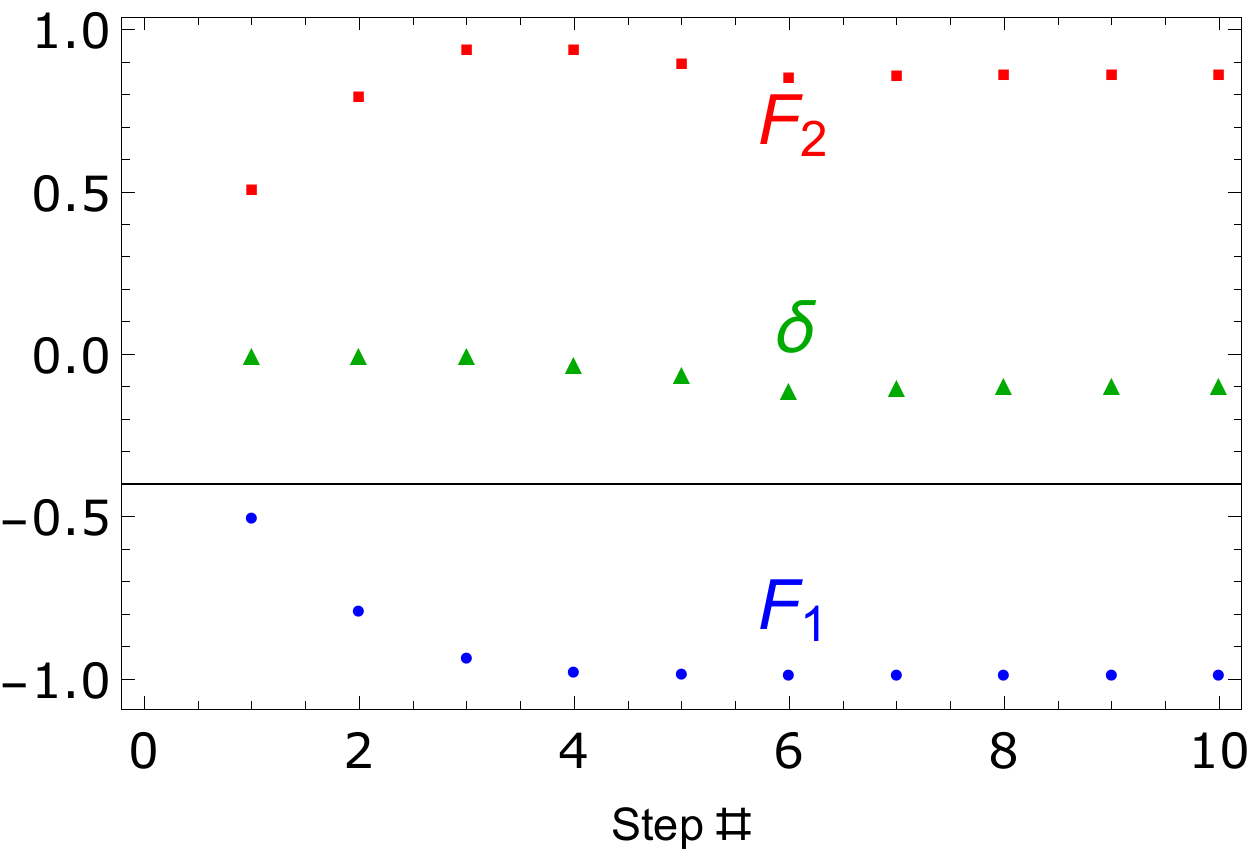}
   \caption{Left, the values of $F_1$ (blue circles) and $F_2$ (red squares) at steps taken by solver before finding the solution. Right, the same but this time an extra variable $\delta$ (green triangles) is added. The solution was found in 10 steps.}
   \label{fig:Translation}
\end{figure}

We now choose $x_1= -5$, $x_2=1$, and $\epsilon=10^{-3}$.  Without
adding an extra variable it took 299 steps for the solver to find
$F_1$ and $F_2$. However, because we know the broken symmetry is
translation, we simply allow for a shift in the values of $x$ by
changing the equations to
\bea
\label{trans3f}
f(x) &=& a e^ {x- s \delta} + \epsilon e^{2(x-s \delta)} - 1~,\\
\label{trans3g}
g(x) &=& 1 - b e^{-(1+\epsilon) (x-s \delta)}~.
\eea
Because we are using a hybrid solver which is sensitive to the scaling
of variables, we shifted by $s \delta$ with $s=10$ to encourage the
solver to make use of $\delta$. Now there will be an infinite set of
solutions $\{F_1, F_2, \delta\}$. After finding one such solution, we
recover the original values of $F_1$ and $F_2$, by evaluating
\eqref{trans3f} at $x_1 + s \delta$ and \eqref{trans3g} at $x_2 + s
\delta$.  With $\delta$ added, it took only 10 steps for the solver
to find the solution. We show the steps that the solver took for this
problem in Fig.~\ref{fig:Translation}.

In the next section we present an important physics problem which we
solved in much more generality in \cite{Masoumi:2016wot}~.
\section{Tunneling in field theories }\label{sec:tunneling}
\begin{figure}
   \centering
   \includegraphics[width=2.5in]{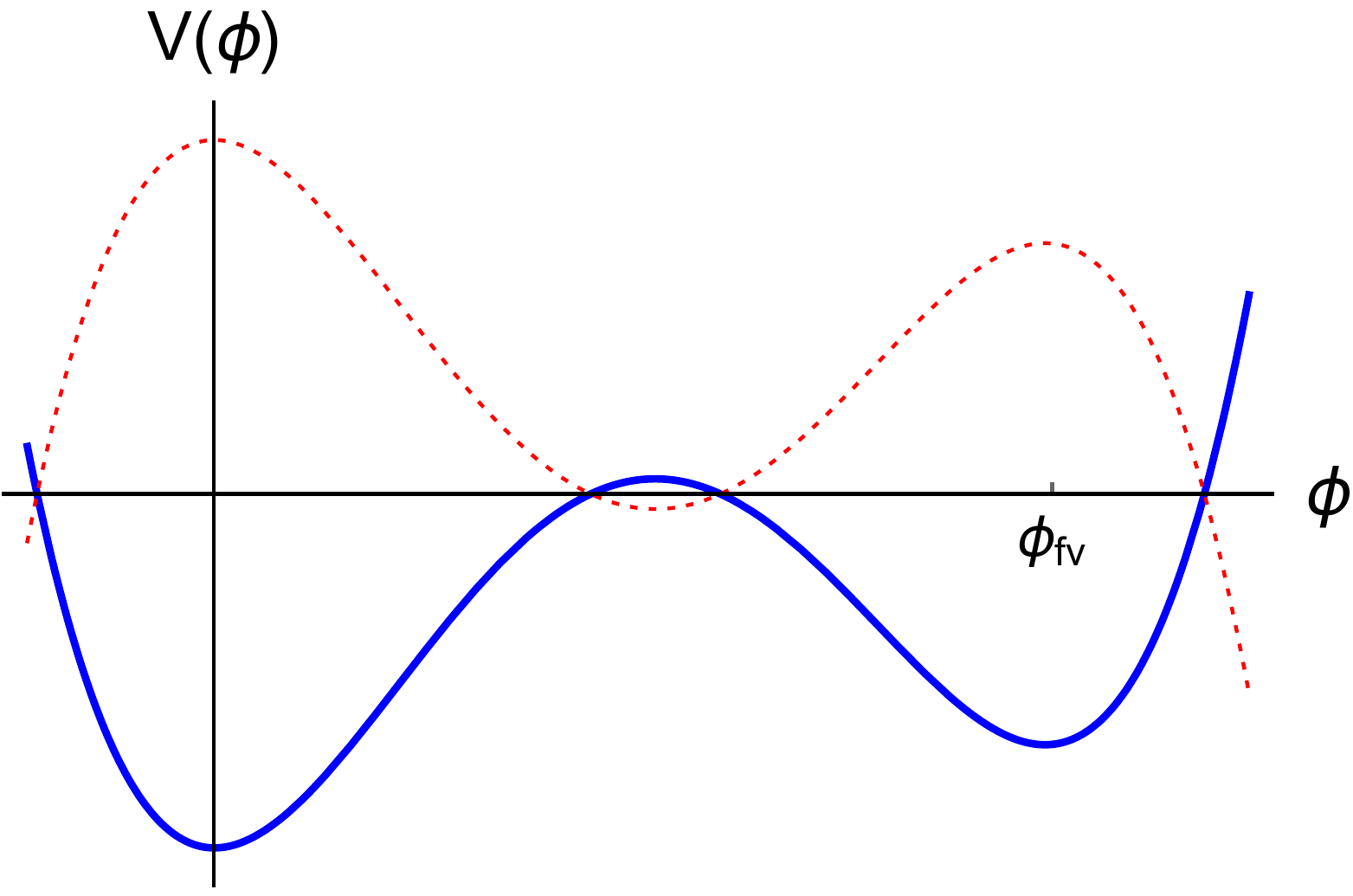} 
   \includegraphics[width=2.5in]{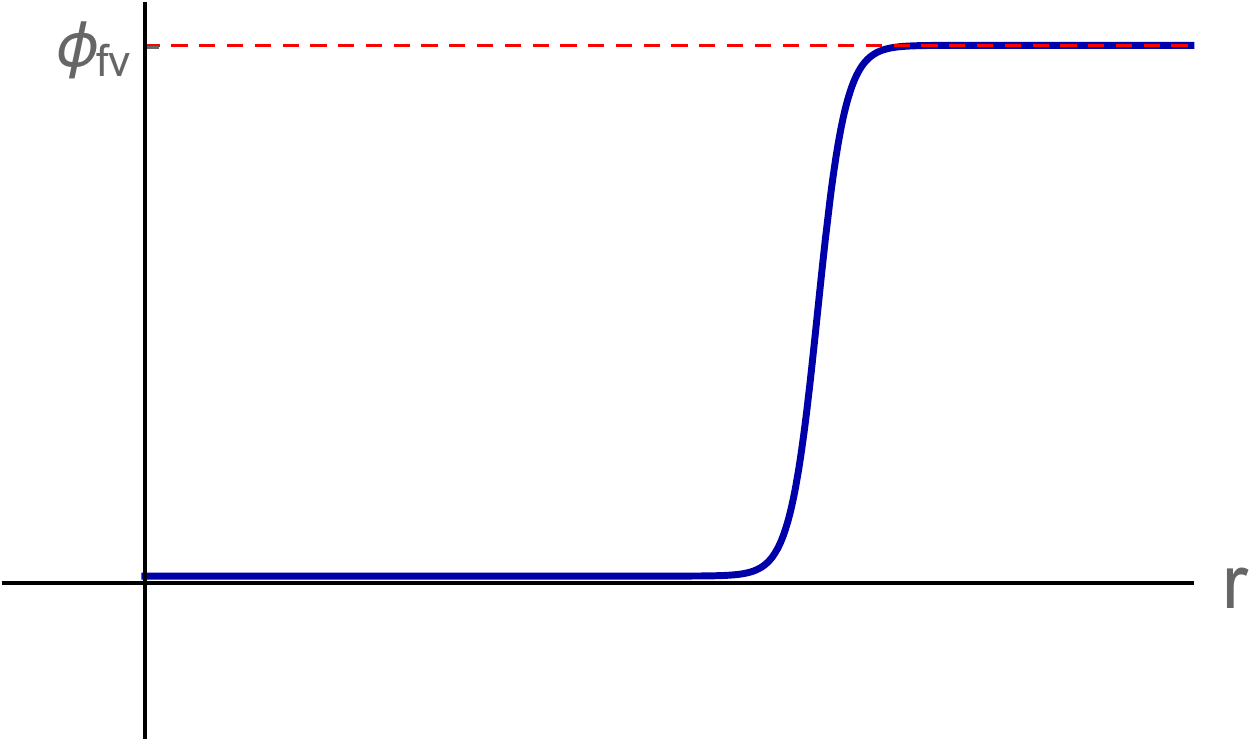} 
   \caption{Left panel, a potential with two minima at $\phi_t=0$ and $\phi_f$ in solid blue and the upside down potential as the dotted red graph.. The minimum at $\phi_f$ is a metastable minimum and can tunnel quantum mechanically to the other minimum. Right panel, the solution to the field equation.}
   \label{fig:pot}
\end{figure}

Problems of differential equations with boundary conditions at a specific point are commonplace in physics. One important examples is the equations used for calculation of cosmological phase transitions. Here we explain the problem briefly. The details can be found in \cite{Coleman:1977py}.  In Fig.~\ref{fig:pot} we present a potential which has a metastable (false) minimum at $\phi_f$ and  a stable (true) minimum at $\phi_t=0$. Finding the lifetime of this metastable minimum  is tantamount to finding the solution of the differential equation 
\bel{EOMs}
	\phi''(r)+ \frac3r \phi'(r)= \frac{\partial U}{\partial \phi}~,
\ee
with two boundary conditions,
\bel{BCS}
	\phi'(0)=0~, \qquad \phi(\infty)=\phi_f~.
\ee
This is the same as the motion of a particle in the upside down
potential shown in the red dotted graph in Fig.~\ref{fig:pot}, under
the influence of a velocity dependent friction given by $-3
\phi'(r)/r$. The standard technique is guessing $\phi(0)$ such that
after evolving the fields it approaches $\phi_f$ .  The field
profile that solves this equation is shown in the right panel of
Fig.~\ref{fig:pot}. It is easy to see that this problem always allows
a solution\footnote{This solution is not necessarily unique, as
  explained in \cite{Masoumi:2016wot}.}.  If one chooses the value
$\phi(0)$ very close to $\phi_t$ the field does not have any
significant change until $r$ gets large. But then the friction term is
negligible and the ``energy" is conserved and it passes $\phi_f$. On
the other hand if one chooses $\phi(0)$ such that $U(\phi(0)) <
\phi_f$, the particle can never reach $\phi_f$. The boundary between
these two is the correct solution that asymptotes to $\phi_f$ at
infinite $r$. 

\begin{figure}
   \centering
   \includegraphics[width=3.5in]{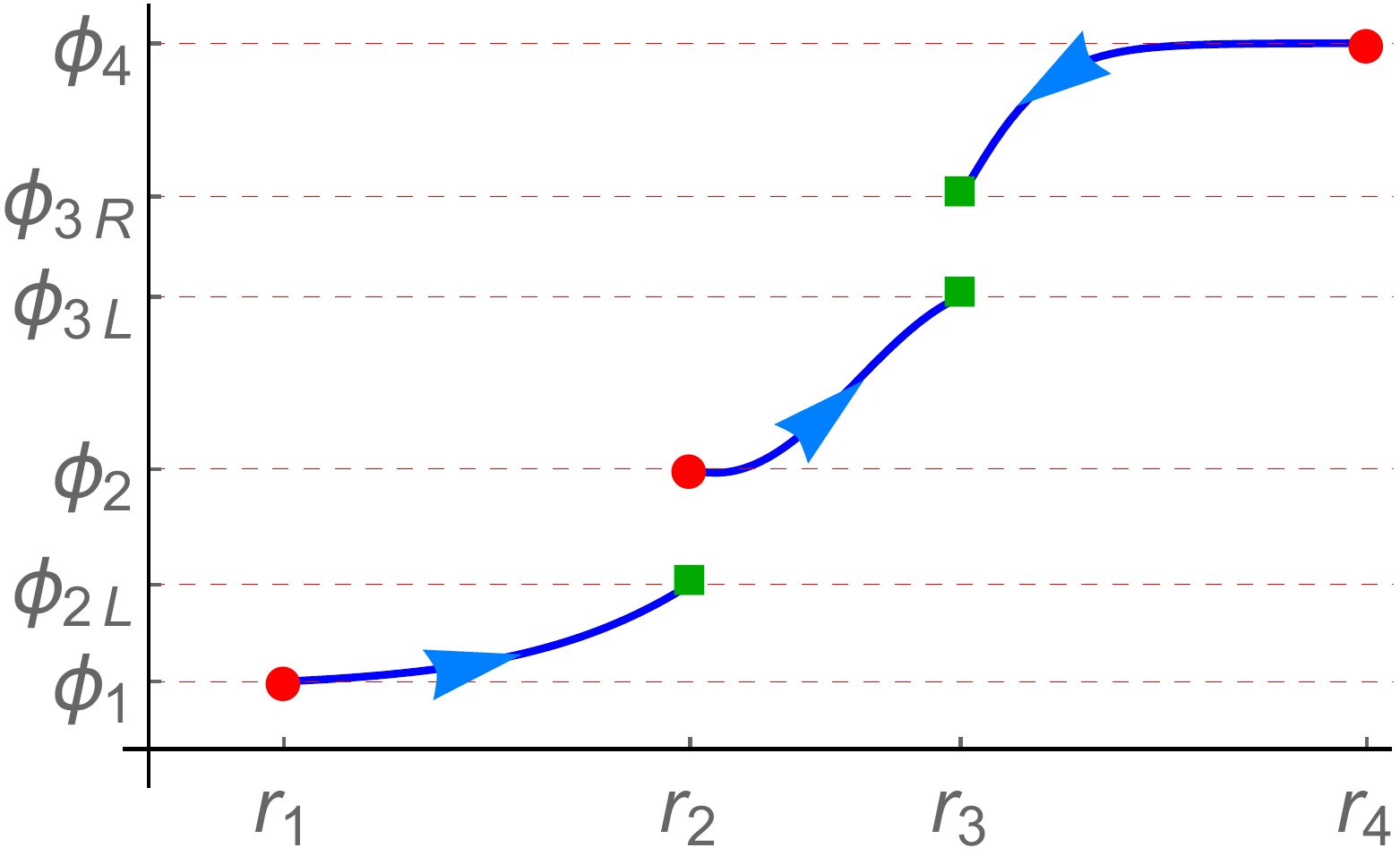} 
   \caption{An illustration of multi-shooting method. One has to adjust the values of $\{\phi_1, \phi_2, \phi'_2, \phi_4\}$ and evolve the field equations \eqref{EOMs} along the blue curves  such that the field and its derivative are continuous at $r_2$ and $r_3$. This produces a system of four equations of four variables.  }
   \label{fig:multiShooting}
\end{figure}

Generalization of this method to more than one field dimension done in
\cite{Masoumi:2016wot} requires a different technique which is called
``multi-interval shooting". This method with 3 intervals is shown in
Fig.~\ref{fig:multiShooting}. One guesses the value of the field at $r_1$
and $r_4$ and the value of the field and its derivative at $r_2$. We
denoted these as $\{\phi_1, \phi_2, \phi'_2, \phi_4\}$. Having these
and using analytic solutions near the true and false minima one can
evolve the field equation \eqref{EOMs} along the blue curve. The goal
is finding the correct values of $\{\phi_1, \phi_2, \phi'_2, \phi_4\}$
such that the field and its derivative are continuous at $r_2$ and
$r_3$. This creates a system of four equation of four variables. If 
not for the middle (friction) term in \eqref{EOMs} and
the fact that $\phi(0)$ is small but nonzero, it would possess a
translational symmetry. Neglecting these two, if $\phi(r)$ satisfied
the equations of motion and the boundary conditions, $\phi(r+\delta)$
would satisfy the same equations.

\begin{figure}
   \centering
   \includegraphics[width=3.3in]{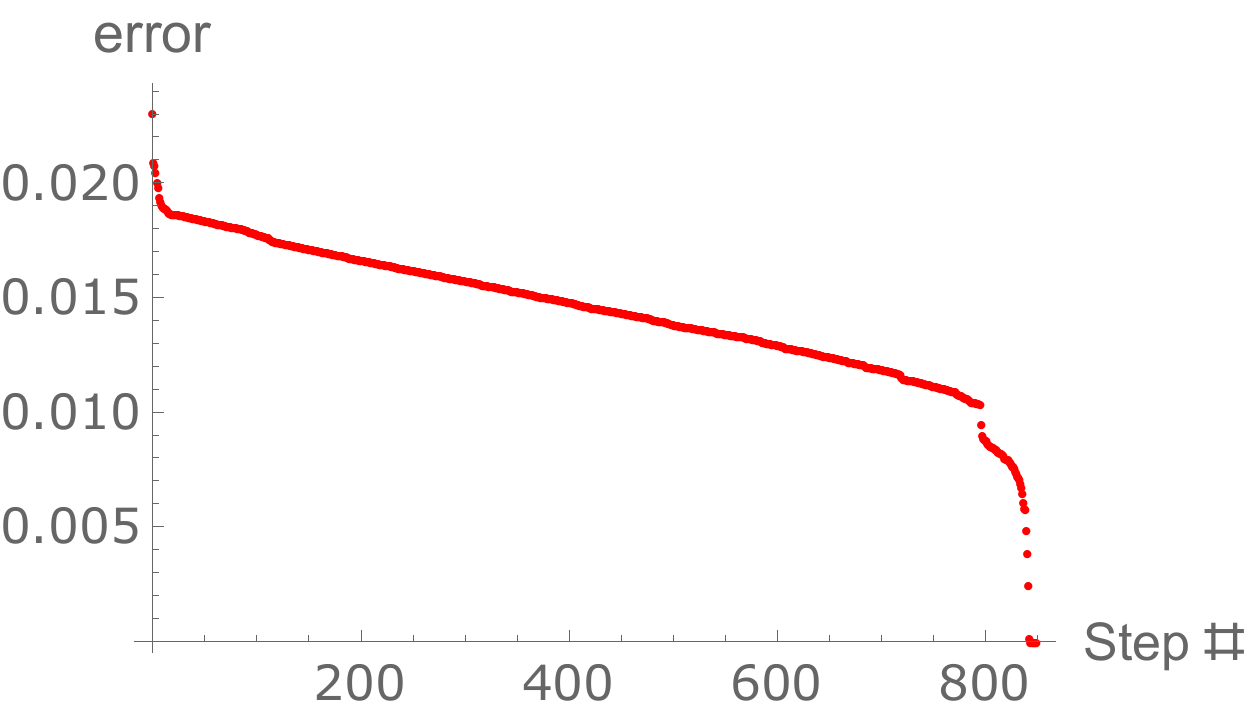} 
   \includegraphics[width=3.3in]{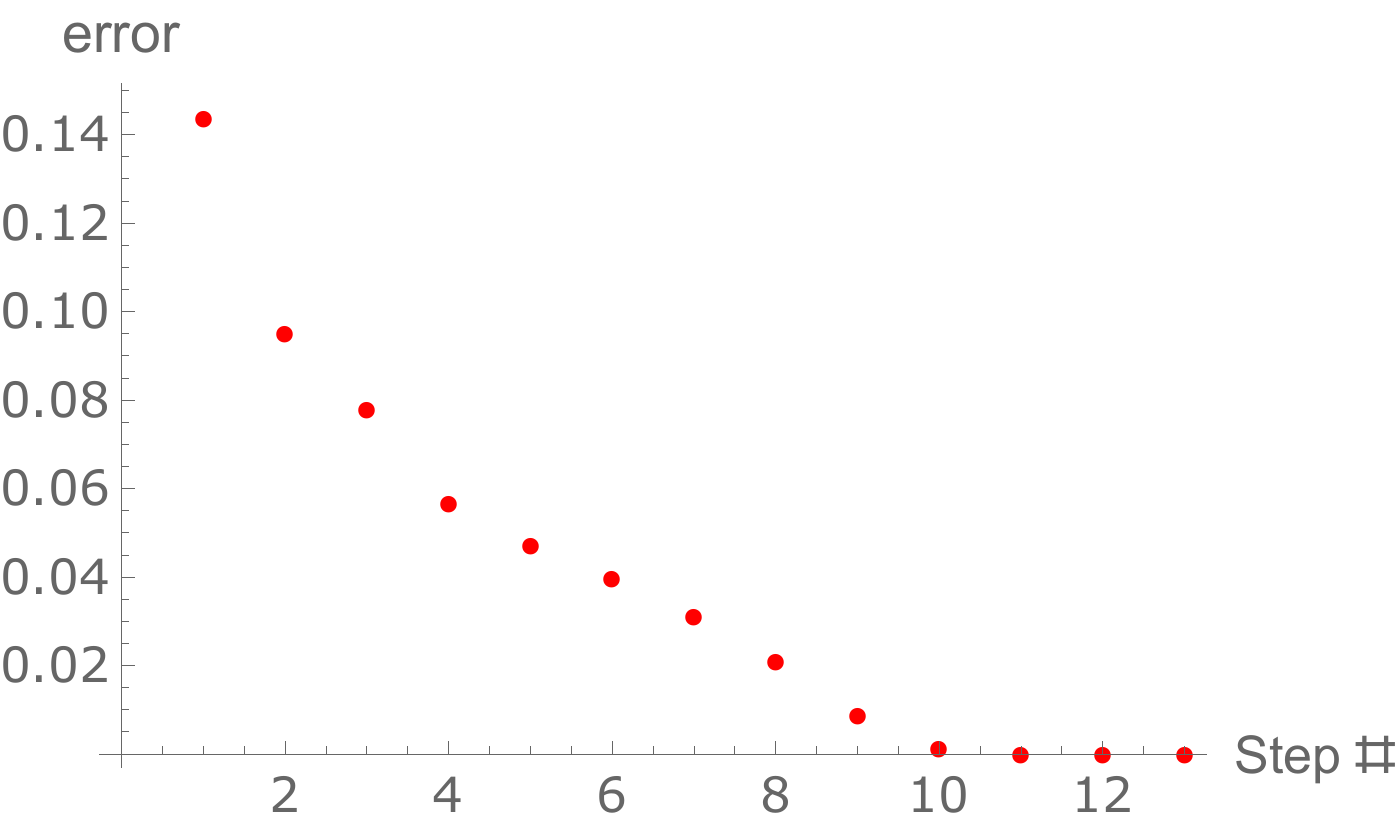} 
   \caption{Left, the change of error after each step taken without adding an auxiliary variable and right the same when including $\delta$ for the potential in \eqref{simplePot} in order to solve \eqref{EOMs}. }
   \label{fig:profileError}
\end{figure}

For the simple potential
\bel{simplePot} U(\phi)=0.2 \phi - 2 \phi^2 + \phi^4~, \quad
\phi_t\approx-1.024~, \quad \phi_f\approx0.9740~,
\ee
 it took the hybrid solver 851 steps to find the solution. We chose the values of $\{r_1, r_2, r_3,
r_4\}=\{12.82,14.03,15.23,16.43\}$ using an analogy with the thin-wall solution\footnote{The details can be found in \cite{Masoumi:2016wot}}. Because the softly broken
symmetry is translation, when we specified the values of the field and
derivative at $\{r_1+\delta, r_2+\delta, r_3+\delta, r_4+\delta\}$
instead of $\{r_1, r_2, r_3, r_4\}$, allowing the solver to exploit
the symmetry, the hybrid solver could find the solution in 12
steps. Most of the steps were taken in the ``valley" which
corresponds to a rigid shift of the profile $\phi(r)$. This is
similar to the previous cases where the solver does not make much
progress in each step as it moving along the ``valley". In
Fig.~\ref{fig:profileError} we show the change of error with and
without adding the auxiliary variable $\delta$.

\section{Powell-Hybrid package} \label{sec:package}

We include with this paper a Mathematica code developed by one of us
(K.D.O.), for the solution of simultaneous nonlinear equations using
Powell's hybrid method \cite{Powell1}.  This code include the
extension described here for solving $N+K$ equations in $N$
variables.  However, it handles only problems where the Jacobian
can be computed analytically.  Powell \cite{Powell1} also includes a
method for approximating the Jacobian using, mainly, the successive
function evaluations, but we did not implement that.

The code and a manual for using it can be downloaded from\break
\texttt{http://cosmos.phy.tufts.edu/Powell}.  This code is the same one
used in \cite{Masoumi:2016wot}.

\section{Conclusion}\label{sec:Conclusion}
In this paper we describe a new method for solving equations with a
softly broken symmetry. The broken symmetry makes it very difficult
for the solver to make progress towards minimizing the error in the
desired equations. The error as a function of the variables usually
has narrow valleys where moving along these valleys does not make
large improvement. As a result the steps that the solver takes are
small and it takes a very large number of steps for the solver to
converge on the solution.

We introduced a method where one adds a set of auxiliary variables
which are the generators of the softly broken symmetry. Adding these
variables makes the solver take large steps in the direction along the
valley and hence converges very quickly. Our method may be applicable
beyond the theories with softly broken symmetries. We believe whenever
such a valley is present adding variables which make the solver move
in the proper direction makes converging much faster and the basin of
attraction larger. However, for general valleys which are not the
results of broken symmetries it may not be easy to identify the
correct auxiliary variables to be added. For this reason we only
mentioned these cases which we know the auxiliary variables must
generate the broken symmetry. These techniques will be much more
powerful and versatile if one can find a systematic way to determine
the needed auxiliary variables.

\section{Acknowledgement}
\label{sec:Acknowledgement}
This work was supported in part by the National Science Foundation
[grant number PHY-1518742].

\bibliographystyle{elsarticle-num}
\bibliography{reference}

\begin{thebibliography}{1}

\bibitem{NumericalRecipe}
William~H. Press, Saul~A. Teukolsky, William~T. Vetterling, and Brian~P.
  Flannery.
\newblock {\em Numerical Recipes 3rd Edition: The Art of Scientific Computing}.
\newblock Cambridge University Press, New York, NY, USA, 3 edition, 2007.

\bibitem{Powell1}
M.~J.~D. Powell.
\newblock A hybrid method for nonlinear equations.
\newblock In Philip Rabinowitz, editor, {\em Numerical methods for nonlinear
  algebraic equations}, chapter~6, pages 87--114. Gordon and Breach Science
  Publishers, New York, 1970.

\bibitem{Masoumi:2016wot}
Ali Masoumi, Ken~D. Olum, and Benjamin Shlaer.
\newblock {Efficient numerical solution to vacuum decay with many fields}.
\newblock 2016.

\bibitem{Coleman:1977py}
Sidney~R. Coleman.
\newblock {The Fate of the False Vacuum. 1. Semiclassical Theory}.
\newblock {\em Phys.Rev.}, D15:2929--2936, 1977.

\end{thebibliography}

\end{document}